\documentclass[english,twocolumn,prl]{revtex4}
\usepackage[T1]{fontenc}
\usepackage[latin9]{inputenc}
\usepackage{babel}
\usepackage{amsmath}
\usepackage{amssymb}
\usepackage{graphicx}
\usepackage{esint}
\usepackage[unicode=true,
 bookmarks=true,bookmarksnumbered=true,bookmarksopen=false,
 breaklinks=false,pdfborder={0 0 1},backref=false,colorlinks=false]
 {hyperref}

\makeatletter

\@ifundefined{textcolor}{}
{%
 \definecolor{BLACK}{gray}{0}
 \definecolor{WHITE}{gray}{1}
 \definecolor{RED}{rgb}{1,0,0}
 \definecolor{GREEN}{rgb}{0,1,0}
 \definecolor{BLUE}{rgb}{0,0,1}
 \definecolor{CYAN}{cmyk}{1,0,0,0}
 \definecolor{MAGENTA}{cmyk}{0,1,0,0}
 \definecolor{YELLOW}{cmyk}{0,0,1,0}
}

\usepackage{braket}
\usepackage{times}
\usepackage{color}

\makeatother

\begin{document}

\title{Equilibrium states of generic quantum systems subject to  periodic driving}

\author{Achilleas Lazarides$^{1}$, Arnab Das$^{2}$ and Roderich Moessner$^{1}$}
\affiliation{$^{1}$ Max-Planck-Institut f\"ur Physik komplexer Systeme, 01187 Dresden, Germany}
\affiliation{$^{2}$ Theoretical Physics Department, Indian Association for the Cultivation of Science, Kolkata 700032, India}

\begin{abstract}
When a closed quantum system is driven periodically with period $T$, 
it approaches a periodic state synchronized with
the drive in which any local observable measured stroboscopically approaches
a steady value.  For integrable systems, the resulting behaviour is
captured by a periodic version of a generalized Gibbs ensemble. 
By contrast, here we show that for generic
non-integrable interacting systems, local observables
become \emph{independent of the initial state} entirely. Essentially,
this happens because Floquet eigenstates of the driven system at
quasienergy $\omega_\alpha$ consist of a mixture of the 
{\em exponentially many} eigenstates of the undriven Hamiltonian 
which are thus drawn from the {\em entire} extensive undriven spectrum. This is a form of
equilibration which depends only on the Hilbert space of
the undriven system and not on any details of its Hamiltonian. 
\end{abstract}
\maketitle

\emph{Introduction\textendash}There has been intense recent interest in
 equilibration and thermalization of closed quantum systems. 
If large enough, such systems approach a steady state
well-described by the usual constructs of statistical mechanics.
The effort to understand the mechanisms by which unitary quantum evolution
leads to time-independent states which can be characterised by fixing
a reasonably small number of observables, as it must if statistical
mechanics is to apply, has been one of the most fruitful in nonequilibrium
quantum dynamics~\cite{srednicki1994chaos,PhysRevA.43.2046,rigol452thermalization,
	Popescu:NaturePhysics:2006,PhysRevE.79.061103,PhysRevLett.101.190403,
	Reimann:NewJournalOfPhysics:2012}. 

At the same time, much experimental and theoretical effort has been
devoted to periodically-driven systems~\cite{grifoni1998driven}. The formal framework
has been mostly set up by Shirley \cite{shirley1965solution} and
Sambe \cite{PhysRevA.7.2203}, and has been successfully applied in
various fields, such as NMR \cite{Leskes2010345,PhysRevB.25.6622},
nonlinear optics~\cite{Chu89} and others~\cite{Eckardt:2005p10775,Lim:2008p422,
	PhysRevB.82.172402}. Closer to the subject of this work, 
it has recently been shown that isolated many-body periodically-driven systems
eventually synchronize into a periodic steady state with the driving 
\cite{Russomanno:PhysicalReviewLetters:2012,Lazarides2014a},
in analogy with closed, non-driven systems approaching a stationary 
equilibrium state.

In a recent article \cite{Lazarides2014a}, we have taken a first step
towards characterising the long-time synchronised state, by obtaining
a description of the long-time steady-state of an integrable system
analogous to the generalized Gibbs ensemble (GGE) \cite{rigol2007relaxation}
for undriven systems, finding that memory of the relevant conserved
quantities persists for all time. 

Here we study the generic situation of a nonintegrable periodically
driven model.  Remarkably, we find that the long-time behaviour is
stationary and independent of both the initial condition and details
of the undriven Hamiltonian beyond its Hilbert space. 

We give a physical mechanism explaining this result: the expectation
values of observables in any eigenstate are the same for all
eigenstates. This is caused by the width of the quasienergy spectrum
being finite, whereas that of the energy spectrum of the undriven
Hamiltonian is extensive. This leads to a perturbation theory in the
driving having vanishing radius of convergence, instead immediately
mixing any initial state with a finite fraction of the states of the
entire spectrum in the thermodynamic limit.

The importance of this feature of the quasienergy spectrum appears to
have been appreciated first by Hone, Ketzmerick and
Kohn in the context of continuum single-particle
problems~\cite{hone1997time,hone2009statistical}. Our
result is also in keeping with a very recent preprint of D'Alessio and Rigol, who
argued that closed driven quantum systems tend to a circular ensemble
of random matrix theory, which they interpret as an infinite temperature
state~\cite{DAlessio2014a}.

The remainder of this paper is organized as follows. We first define the problem
and introduce notation, before deriving our central result of the existence of a
steady state which is independent of all of time, driving, and the undriven Hamiltonian,
depending only on the Hilbert space. We then analyze  a particular model
Hamiltonian numerically in order to demonstrate the correctness of the 
central ingredients of our analysis. 
We then discuss the difference to integrable systems, and conclude with an outlook.

\emph{Setup\textendash{}}We consider a periodically-driven system
described by the a Hamiltonian
\begin{equation}
H(t)=H_{S}+uh_{D}(t)\label{eq:H-general}
\end{equation}
with $H_{S}$ time-independent and nonintegrable and $h_{D}(t+T)=h_{D}(t)$
the periodically-driven part, with $u$ a driving amplitude with units
of energy. 

We shall take $H_{S}$ to satisfy the eigenstate thermalization hypothesis
(ETH):  eigenstates that are close in energy look ``similar''~\cite{pechukas1983distribution}; 
this notion has been made more concrete
recently~\cite{PhysRevA.43.2046,srednicki1994chaos,rigol452thermalization,
	beugeling2013finite}.
Following these, we say that the ETH is satisfied for a certain
operator $\hat{O}$ if the eigenstate expectation values (EEVs) of
$\hat{O}$, defined as $\bra{\varepsilon_{\alpha}}\hat{{O}}\ket{\varepsilon_{\alpha}}$
with $\ket\varepsilon$ an energy eigenstate of energy $\varepsilon$ form
a smooth function of the eigenstate energy $\varepsilon$ in the thermodynamic
limit. Thus the mean energy of
a macroscopic system fixes the expectation value of $\hat{O}$, and a
small variation in the energy results in a small variation in the
expectation value. Had this not been the case, arbitrarily small (microscopic)
changes in energy would result in vastly different expectation values
of the operator on macroscopic scales. The ETH has been confirmed
to occur in a number of systems~\cite{beugeling2013finite,rigol452thermalization,
	PhysRevE.85.060101}

Adding a periodically-driven part, $uh_{D}(t)$, to $H_{S}$
necessitates a change of viewpoint: instead of eigenstates and
eigenenergies, one considers Floquet states and quasienergies. These
are obtained from the eigenfunctions and eigenvalues of the operator
\begin{equation}
U\left(\epsilon,\epsilon+T\right)=\mathcal{T}\exp\left(-i\int_{\epsilon}^{\epsilon+T}dt\: H(t)\right)\label{eq:u-period}
\end{equation}
form which one may define an effective Hamiltonian via
$\exp\left(-iH_{eff}\left(\epsilon\right)T\right)=U\left(\epsilon,\epsilon+T\right)$.
Its eigenvectors $\ket{\alpha(\epsilon)}$ 
satisfy $\ket{\alpha(\epsilon)}=\ket{\alpha(\epsilon+T)}$,
while its eigenvalues have the form $\exp\left(-i\omega_{\alpha}T\right)$,
with \emph{quasienergies }$\omega_{\alpha}$  independent of
$\epsilon$. The Floquet states, forming a complete set for equal-time
properties, are defined as $\ket{u_{\alpha}(t)}=\exp\left(-i\omega_{\alpha}t\right)\ket{\alpha(t)}$.

Note that, as behooves a periodic system, the quasienergies necessarily 
lie in a "Brillouin zone" (BZ) of finite, non-extensive
width $\omega=2\pi/T$; this feature will play a crucial role in our analysis. 

Starting from an initial state, after a transient period synchronization
with the driving is achieved~\cite{Russomanno:PhysicalReviewLetters:2012,Lazarides2014a}
in the following sense: Take an initial density matrix 
$\hat{\rho}(0)=\sum_{\alpha,\beta}\rho_{\alpha,\beta}\ket{\alpha(0)}\bra{\beta(0)}$;
at long times, the system behaves indistinguishably~\cite{PhysRevLett.101.190403,
	Reimann:NewJournalOfPhysics:2012}
from one described by 
\begin{equation}
\hat{\rho}_{DE}(t)=\sum_{\alpha}\rho_{\alpha,\alpha}\ket{\alpha(t)}\bra{\alpha(t)},\label{eq:synced-rho}
\end{equation}
which is evidently periodic in time. The expectation value of an operator
$\hat{O}$ in this state is
\begin{equation}
\mathcal{O}(t)=\sum_{\alpha}\rho_{\alpha,\alpha}O_{\alpha,\alpha}(t)\label{eq:synced-O}
\end{equation}
with the eigenstate expectation values (EEVs) $O_{\alpha,\alpha}(t)=\bra{\alpha(t)}\hat{O}\ket{\alpha(t)}$.
This is analogous to the so-called Diagonal Ensemble (DE) for non-driven
systems; in principle, this depends on the initial state through the
quantities $\rho_{\alpha,\alpha}$.

\emph{Eigenstate mixing\textendash{}}We begin by discussing the eigenvectors
and quasinergies of $H_{eff}(\epsilon)$. For vanishing $u$ (alternatively,
for stroboscopic observations of the system in the absence of any
driving), the eigenstates $\ket{\alpha\left(\epsilon\right)}$ are
time-independent and identical to those of $H_{S}$.
The corresponding quasienergies are thus 
$\omega_{\alpha}=\mod\left(\varepsilon_{\alpha},\omega\right)$.
This implies that, even if the (non-driven) system $H_{S}$ satisfies
the ETH for some observable, labelling the eigenstates by $\omega_{\alpha}$
instead of $\varepsilon_{\alpha}$ will in general destroy this property: the EEVs
will not be a smooth function of the quasienergy, since now eigenstates
whose energy differs by an integer number of  'reciprocal lattice
vectors' $\omega=h/T$ (with $h$ Planck's constant) have the same
quasienergy. By continuity, one might expect this to remain true
for ``small'' $u$. However, this expectation turns out to be wrong
in the thermodynamic limit.

One may see this from the results of Ref. \cite{hone1997time,hone2009statistical},
from which the following picture emerges: Suppose we fix a $u$ and
calculate the states $\ket{\alpha\left(\epsilon\right)}$. If $u\rightarrow u+\delta u$,
one might hope to use perturbation theory to obtain the new states.
However, the quantity compared to which $\delta u$ needs to be small is
the quasienergy level spacing. As the dimension of Hilbert $D_{H}$ increases
exponentially with system size, and the width of the quasienergy BZ is independent of it,
the level spacing is exponentially small, and hence so is  
the radius of convergence of such a perturbation theory --  
one cannot expect adiabatic evolution.  
In particular, the basis states at arbitrarily small $u$
are not perturbatively related to the undriven ones in the thermodynamic
limit; and an arbitrarily small change in $u$
mixes the $\ket{\alpha\left(\epsilon\right)}$ among
themselves.

To be more precise, the condition  
$u \overline{h}_D/\hbar\omega\gg 1$ where $\overline{h}_D$ the typical
magnitude of a matrix element of $h_D$ ensures that each state is coupled to
the entire Brillouin zone, and is sufficient. 
The magnitude of $\overline{h}_D$ for a global periodic driving term
grows as a power of the system size. This condition
is satisfied in the thermodynamic limit, and an arbitrarily small $u$ should suffice
for our results to hold.

Applying these results to a $u$ satisfying the above conditions,
we see that each $\ket{\alpha\left(\epsilon\right)}$ contains contributions
from bands of width set by $u\overline{h}_D$ uniformly spread over the entire undriven
spectrum, so that a finite fraction of the undriven eigenstates participate.
We have confirmed this explicitly by calculating the
average participation ratio of the eigenstates of $H_{eff}\left(\epsilon\right)$ %
in the basis of the eigenstates of $H_{S}$ (see Supplementary Material).

\begin{figure}[t]
\includegraphics[scale=0.75]{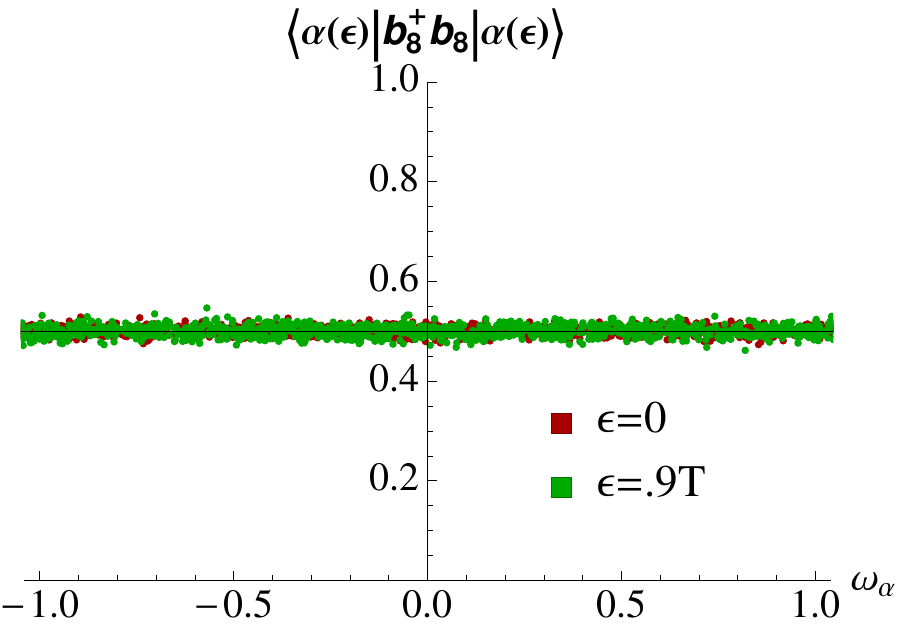}

\caption{Example of the EEV dependence on the quasienergy $\omega_{f}$ for
$u/\hbar\omega=1$ and system size and particle number $L=14$, $N=7$
for a Hilbert space dimension $D_{H}=3432$, with parameters
$u=V_1=V_2=J$ and driving frequency $\hbar \omega=h/T=J/4$. Points indicate
expectation value of the density at site $i=8$ in 
an eigenstate $\ket{\alpha(\epsilon)}$ of $H_{eff}(\epsilon)$
versus the state's quasienergy $\omega_\alpha$ at two different times.
The black line indicates $\mathrm{tr}\left(b^\dagger_8 b_8\right)=N/L=0.5$
\label{fig:eev-example}}
\end{figure}

Given  this strong mixing across the entire spectrum of $H_{S}$, it is thus natural to 
expect that expectation values with respect to the 
$\ket{\alpha\left(\epsilon\right)}$ effectively average essentially uniformly over
those with respect to  the eigenstates of
$H_{S}$, as captured qualitatively by the following rough 
argument. Let us expand the former in terms of the latter, 
$\ket{\alpha\left(\epsilon\right)}=\sum_{n}\ket{n}\bra{n}\left.\alpha\left(\epsilon\right)\right>$,
and replace $\left<\alpha\left(\epsilon\right)\right|\left.m\right>\approx1/\sqrt{D_{H}}\exp\left(i\phi_{m}\left(\epsilon\right)\right)$
with the phases $\phi_{m}(\epsilon)$ uncorrelated between different
$m$. Note that for this replacement to be reasonable, 
provided a smooth dependence of $\bra{m}\hat{O}\ket{m}$ on 
$\varepsilon_m$, it is not necessary
that all overlaps $\left<\alpha\left(\epsilon\right)\right|\left.m\right>$
are finite; rather, a sufficient condition is that the ones that are
finite are uniformly and densely spread throughout the 
band of $H_{S}$, which we numerically observe (see Supplementary Material).
Then, $\bra{\alpha\left(\epsilon\right)}\hat{O}\ket{\alpha\left(\epsilon\right)}\approx D_{H}^{-1}\sum_{m,n}\exp\left[i\left(\phi_{n}\left(\epsilon\right)-\phi_{m}\left(\epsilon\right)\right)\right]\bra{m}\hat{O}\ket{n}$.
Finally, given that a) the phases are uncorrelated and b) $\bra{m}\hat{O}\ket{n}$
decreases rapidly with increasing 
$\left|\varepsilon_{m}-\varepsilon_{n}\right|$, as occurs in 
nonintegrable systems, and assuming that none of the $\bra{m}\hat{O}\ket{n}$
grows with $D_H$ (in other words, that the observable is not localised in the
basis of the $\ket{m}$)
we find $\bra{\alpha\left(\epsilon\right)}\hat{O}\ket{\alpha\left(\epsilon\right)}\approx\frac{1}{D_{H}}\sum_{m}\bra{m}\hat{O}\ket{m}=D_{H}^{-1}\mathrm{{tr}\left(\hat{O}\right)}$,
\emph{independent of both $\alpha$ and $\epsilon$.} 

This implies
that the long-time steady-state of the observable is not just periodic,
but in fact even {\em independent of time}. In addition, since Eq.~\eqref{eq:synced-O}
becomes 
\begin{equation*}
\mathcal{O}(t)=O(t)\sum_{\alpha}\rho_{\alpha,\alpha}=D_{H}^{-1}\mathrm{{tr}\left(\hat{O}\right)},
\end{equation*}
so that the long time state is even completely independent of the initial
condition (encoded in $\rho_{\alpha,\alpha}$). 

We now turn to the numerical confirmation of the ingredients of our above analysis
for a particular instance of a driven model system.

\begin{figure}[t]
\includegraphics[scale=0.8]{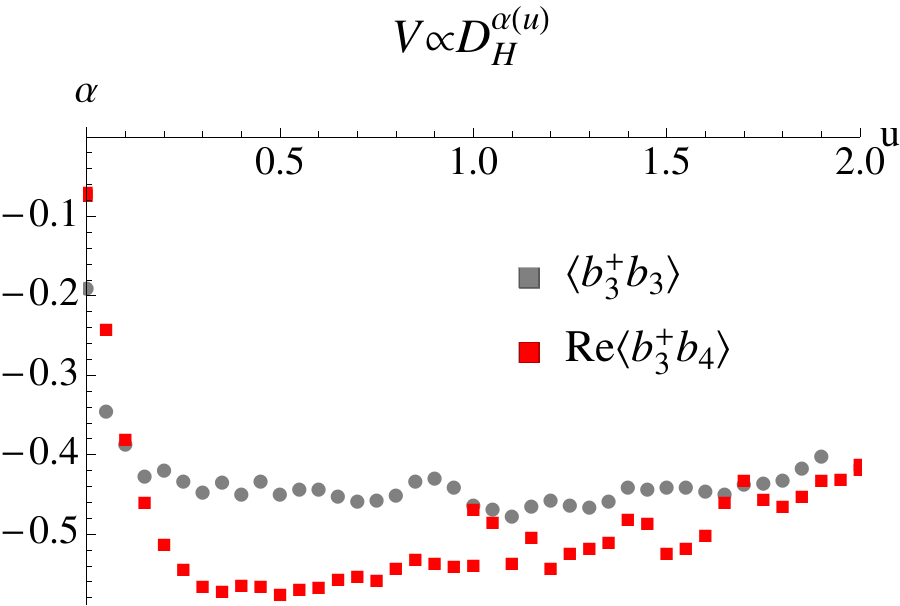}

\caption{Fitted exponent $\alpha$ versus driving amplitude $u$, extracted
for the observable $b_{3}^{\dagger}b_{4}$ as well as the density
at site $i=3$ for the Hamiltonian of Eq. \ref{eq:H} for driving
period $\omega=2\pi/T=1$. The upturn at small $u$ is a finite-size
effect.\label{fig:alpha-vs-u}}
\end{figure}

\emph{Eigenstate expectation values\textendash}We
consider hardcore bosons ($b_i^2=0$) with
\begin{widetext}
\begin{equation}
H(t)=-\frac{1}{2}\sum_{i}b_{i}^{\dagger}b_{i+1}+hc+V_{1}\sum_{i}n_{i}n_{i+1}+V_{2}\sum_{i}n_{i}n_{i+2}+u\sum_{i}V_{i}(t)n_{i}\label{eq:H}
\end{equation}
\end{widetext}
featuring a potential $V_{i}(t)=\widetilde{u}\left(t\right)\left(-1\right)^{i}$
with $\widetilde{u}(t)=+1$ for $0<t<T/2$ and $\widetilde{u}(t)=-1$
for $T/2\leq t\leq T$ (see Supp.~Mat. for another example). 
Throughout, $J=V_{1}=V_{2}=1$.

For this system,~\footnote{We also add a very small tilt 
in the form of a potential $\alpha\sum_{i}n_{i}$
with $\alpha/J=10^{-3}$ to break translational invariance and remove
spurious degeneracies.%
} we calculate the EEVs of the (arbitrarily chosen) local density operator
$b_{8}^{\dagger}b_{8}$, plot them and compare to the mean EEV in
a window centred at the current EEV; an example is shown in Fig.~\ref{fig:eev-example}~
\footnote{The averaging window is taken to be 100 states, but our results are
insensitive to this width. As we shall see, the mean is independent
of quasienergy for large enough systems.%
} As expected, the EEVs show no dependence on quasienergy.
Indeed, this result seems to be natural absent a preferred choice of 
origin of the quasienergy BZ.

We next study the approach to the thermodynamic limit. To do this,
we define a root mean square deviation of the EEVs. Taking an average
over a window of $w+1$ states, $\bar{O}_{\alpha}=\frac{1}{w}\sum_{\beta}O_{\beta,\beta}$ with $\beta$ running from $\alpha-w/2$ to $\alpha+w/2$,
the root mean square deviation is $V^{2}=\frac{1}{D_H}\sum_{\alpha}^{D_H}\left(O_{\alpha\alpha}-\bar{O}_{\alpha}\right)^{2}$.
We are interested in whether and how $V$ vanishes with increasing
$D_{H}$. By numerically fitting its behaviour, we find that $V=cD_{H}^{\alpha}$
(see Supplementary Material for an example fit); the exponent $\alpha$
for a number of different $D_{H}$ (which we vary by varying the system
size $L$, and the number of particles $N$) and two observables,
the density at site $i=8$ and the operator $b_{3}^{\dagger}b_{4}$,
is shown in Fig. \ref{fig:alpha-vs-u}.

From this, $\alpha$ appears to be independent of $u$ and approximately equal
to $-1/2$; the upwards shift for small $u$ is a finite-size effect, as $u$ becomes too 
small given the level spacing of the system sizes we have access to. 
We therefore conclude that for a large enough system, the EEVs
$O_{\alpha,\alpha}$ become independent of $\alpha$ in the thermodynamic
limit as expected.

\begin{figure*}[t]
\includegraphics[scale=0.65]{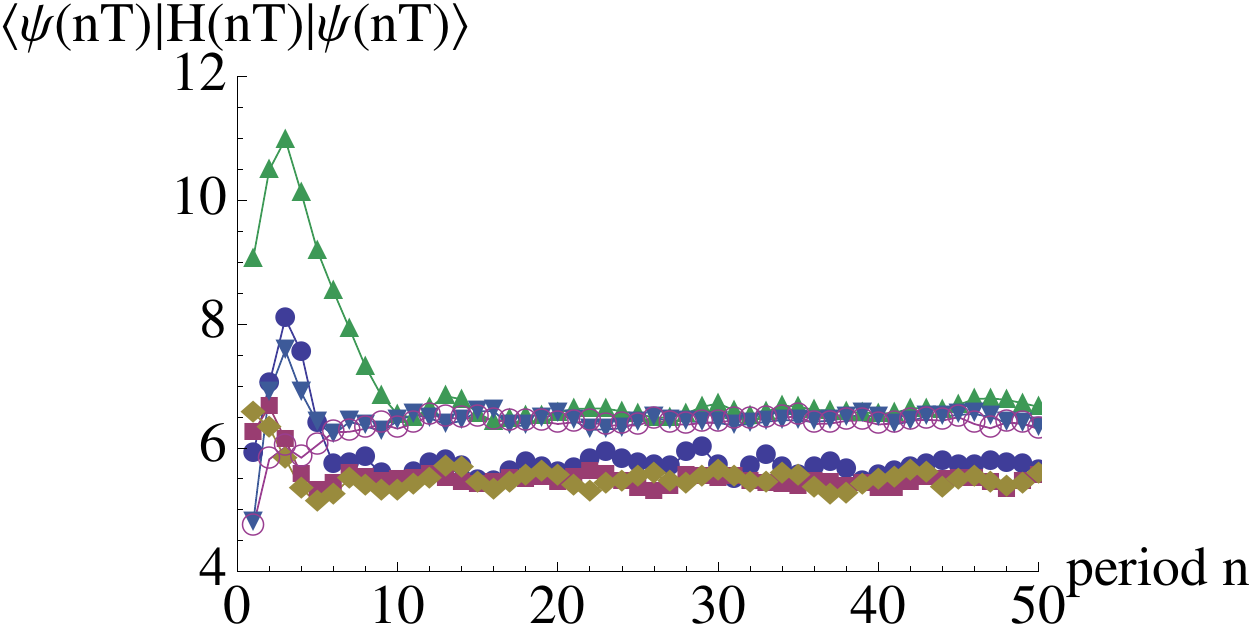}
\includegraphics[scale=0.65]{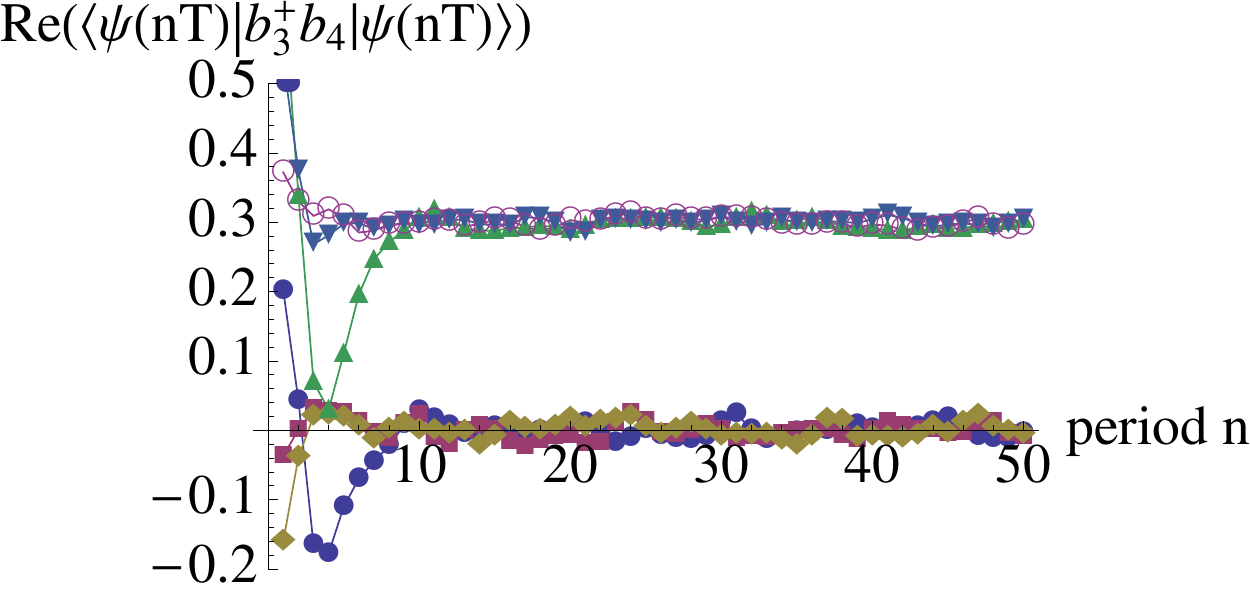}

\caption{\textbf{Left}: Dynamical evolution of instantaneous value of the energy
at the beginning of each period. The blue, gold red and gold data
points (three lower lines) correspond to different states, selected
from different parts of the band of a Hamiltonian (which is different
from the Hamiltonian used during the driving) and for for $L=12$,
$N=6$ so $D_{H}=924$. The three top lines show time evolution of
the same states, but for for $L=14$, $N=7$ so $D_{H}=3003$. In
both cases, $u/\hbar\omega=5$. \textbf{Right}: Similar results are
obtained for other observables, such as $b_{3}^{\dagger}b_{4}$. The
bottom three lines are for $L=12$, $N=6$ while the top three lines
are for $L=14$, $N=7$, and are offset vertically for clarity (in
reality, they also oscillate about 0). \label{fig:dynamical-H}}
\end{figure*}
\emph{Dynamics\textendash{}}Having shown that the EEVs are all
equal, we now confirm that this does indeed lead to independence of
the final state from the initial state. To this end we explicitly
calculate the dynamics starting from different initial states and
check whether the final state is the same. We follow the following
protocol. Diagonalising the Hamiltonian of Eq. \ref{eq:H} with $J=V_{1}=V_{2}=1$
and a diagonal potential $V_{i}=i^{2}$, we select three states: the
ground state, the eigenstate $1/4$ of the way up from the ground state
and the state in the middle of the band. We then switch off the diagonal
potential and, for each state, calculate the time evolution under
periodic driving with, again, $J=V_{1}=V_{2}=1$, $u=5J$ and $V_{i}(t)=u\left(t\right)\left(-1\right)^{i}$
with $u(t)=+u$ for $0<t<T/2$ and $u(t)=-u$ for $T/2\leq t\leq T$,
as in the preceding discussion. At the beginning of each period, we
calculate the instantaneous expectation of the Hamiltonian of Eq.
\ref{eq:H}. The results are displayed in figure \ref{fig:dynamical-H}: for
two different system sizes, the expectation value of the instantaneous
energy evolves to the same value in all three states, as do the expectation
values of the operator $b_{3}^{\dagger}b_{4}$.

\emph{Discussion\textendash}Taking a step back, we recognise two
things happening here. Firstly, at long times the system
approaches a steady state (Eq. \ref{eq:synced-O}), which is in principle
periodic in time. Secondly, and more surprisingly, the EEVs are independent
of the quasienergy, which leads to the synchronised state being independent
of the initial condition. It is rather a property of the basic degrees of freedom
of the system only, such as their 
locality and the Hilbert space they span, being essentially independent of any
further 'details' of the Hamiltonian. 
The system
therefore loses all memory of the initial state, unlike the situation
in either non-driven systems undergoing a quench or integrable driven
systems \cite{Lazarides2014a}. 

The necessary ingredient  is the absence of an adiabatic
limit as $u$ is varied for large enough systems
\cite{hone1997time,hone2009statistical}. This causes an arbitrarily
small change in $u$ to mix all eigenstates together; applying this
to $u$ close to the undriven limit $u=0$, we see that the information
contained in the dependence of the EEVs on energy, which determines
the macroscopic properties of the system as a function of its energy,
is completely scrambled. The final state mixes together macroscopic
properties of undriven states at all energies and ends up completely
featureless as a result.

By contrast, for a non-driven system, a finite-strength perturbation
only couples unperturbed eigenstates within a finite fraction of the energy band,
while. As a consequence, the EEVs of any operator in the perturbed basis are
sensitive only to the unperturbed EEVs from nearby energies.
This results in the perturbed EEVs remaining energy dependent and,
in general, continuous. 

The fact that  this does not occur for \emph{integrable} driven systems, 
where a periodic generalized Gibbs ensemble is found instead \cite{Lazarides2014a}, 
seems at odds with the generality of the above arguments. However, note
that the extensive number (proportional to system size $L$)  of conserved
quantities {\em exponentially} reduces the number of states which get
mixed together, as fixing $L$ quantities independently leads to a Hamiltonian matrix
block diagonal with exponentially many uncoupled blocks. In those cases 
where the driving does not couple the different blocks (as it does not 
for systems mappable
to free fermions via a Jordan-Wigner transformation, for instance)
the scrambling of the eigenstates described above happens only inside
each (small) block of size polynomial in $L$.
This is not sufficient to randomize the
eigenvectors, so that $O_{\alpha\alpha}\left(t\right)$ is neither $t$- nor
$\alpha$-independent and cannot be pulled out of the sum on the right
hand side of Eq. \ref{eq:synced-O}. Therefore, the long-time state is sensitive
to the (initial state dependent) form of $\rho_{\alpha,\alpha}$. 

Finally, we note that our results might be inapplicable to systems with infinite
local Hilbert spaces, such as non-hardcore bosonic or continuum systems. This 
follows from the fact that the diagonal ensemble
result for the long-time expectation value of
the instantaneous energy density $\mathrm{tr}\left(H\left(nT\right)\right)/L$ diverges. 

Several avenues for future work immediately suggest themselves. Firstly, 
it will be interesting to study the approach
to the steady state as a function of time and system size. Secondly, 
it will be interesting to study and classify the effect of driving
for systems not obeying ETH in the undriven limit.
Finally, given our analysis was phrased largely perturbatively in $u$, it is not entirely 
clear what happens when $u$ becomes arbitrarily large. 

\emph{Acknowledgments\textendash}A.~L. thanks W.~Beugeling, A.~Eckardt and 
O.~Tieleman for insightful observations throughout the course of this
work. A.~D. thanks A.~Sen for discussions and MPI-PKS for hospitality during the
course of this work.

\onecolumngrid
\appendix
\newpage


\section{Example results for EEV variance versus system size\label{sec:Example-results-for}}

\begin{figure}[b]
\includegraphics[scale=.8]{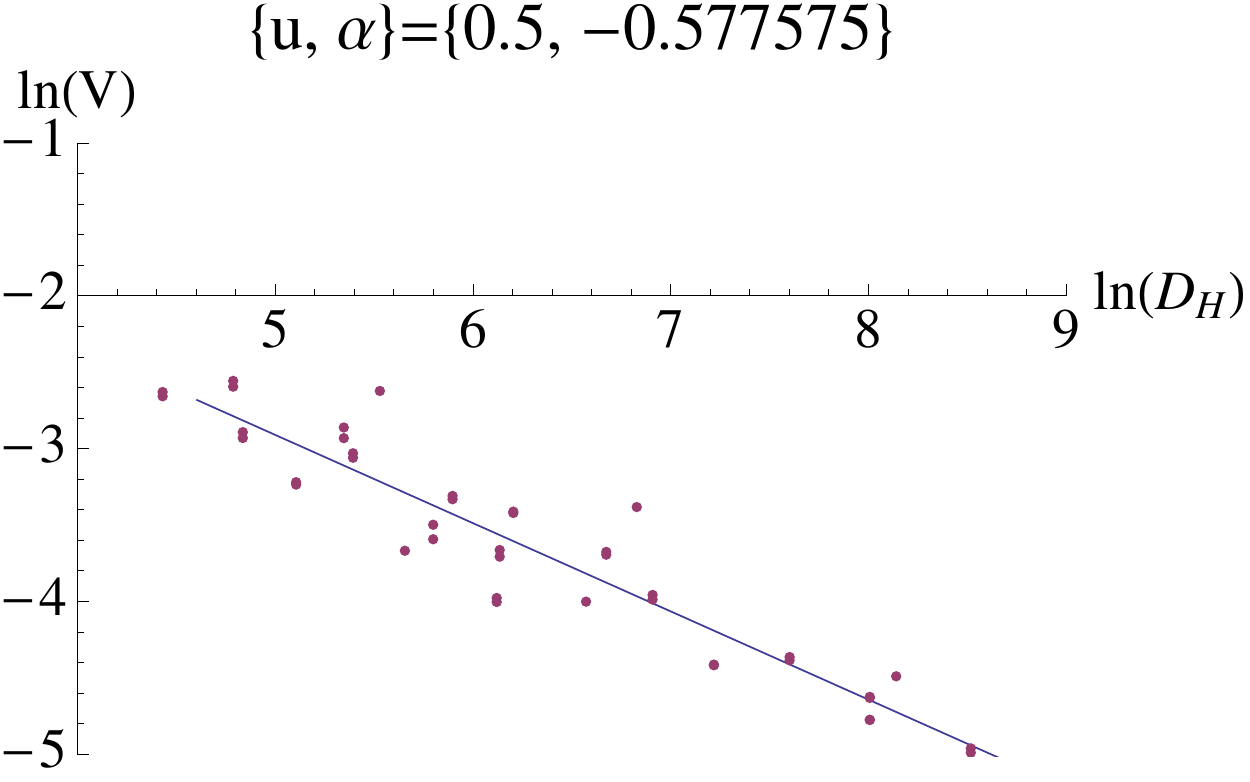}

\caption{Best fit and data points for $u=0.5$. This figure corresponds to
a single data point in Fig.\ref{fig:alpha-vs-u}. It is obtained by
varying $L$, the system size, and $N$, the number of particles.\label{fig:example-fit}}
\end{figure}

Figure \ref{fig:example-fit} shows how the exponent $\alpha$ for
the scaling of the EEV variance $V$ with Hilbert space dimension
$D_{H}$ is extracted. The results shown in Fig. \ref{fig:alpha-vs-u}
of the main text are obtained by repeating this for different values
of $u$.

\section{Participation ratios\label{sec:Participation-ratios}}

Define the participation ratio (PR) $\phi_{\alpha}(\epsilon)=\left(D_{H}\sum_{n=1}^{D_{H}}\left|\left<n\right.\ket{\alpha(\epsilon)}\right|^{4}\right)^{-1}$
and its average $\bar{\phi}\left(\epsilon\right)=D_{H}^{-1}\sum_{\alpha=1}^{D_{H}}\phi_{\alpha}\left(\epsilon\right)$;
the quantity $\phi_{\alpha}\left(\epsilon\right)$ is $1/D_{H}$ if
a single $\ket{n}$ has finite overlap with $\ket{\alpha\left(\epsilon\right)}$
and becomes 1 if every single $\ket{n}$ participates equally in $\ket{\alpha\left(\epsilon\right)}$.
It therefore roughly measures the fraction of the eigenstates of $H_{S}$
mixed into $\ket{\alpha\left(\epsilon\right)}$. For convenience, we
also define $\bar{m}=\frac{1}{\sum_{m}f_{m}}\sum_{m}mf_{m}$ with
$f_{m}=\left|\left<\alpha\left(\epsilon\right)\right.\left|m\right>\right|$
, the mean position of the participating eigenstates, and a ``radius
of gyration'', $w^{2}=\frac{1}{\sum_{m}f_{m}}\sum_{m}\left(m-\bar{m}\right)^{2}f_{m}$.
This roughly indicates how much of the bandwidth of $H_{S}$ is involved
in each eigenstate of $H_{eff}\left(0\right)$. Together, these quantities
allow us to show two things: First, that a \emph{finite fraction}
of eigenstates of $H_{S}$ participate in each $\ket{\alpha(0)}$.
Second, that the \emph{entire bandwidth} of $H_{S}$ participates
in each $\ket{\alpha(0)}.$

\begin{figure}
\includegraphics[scale=.8]{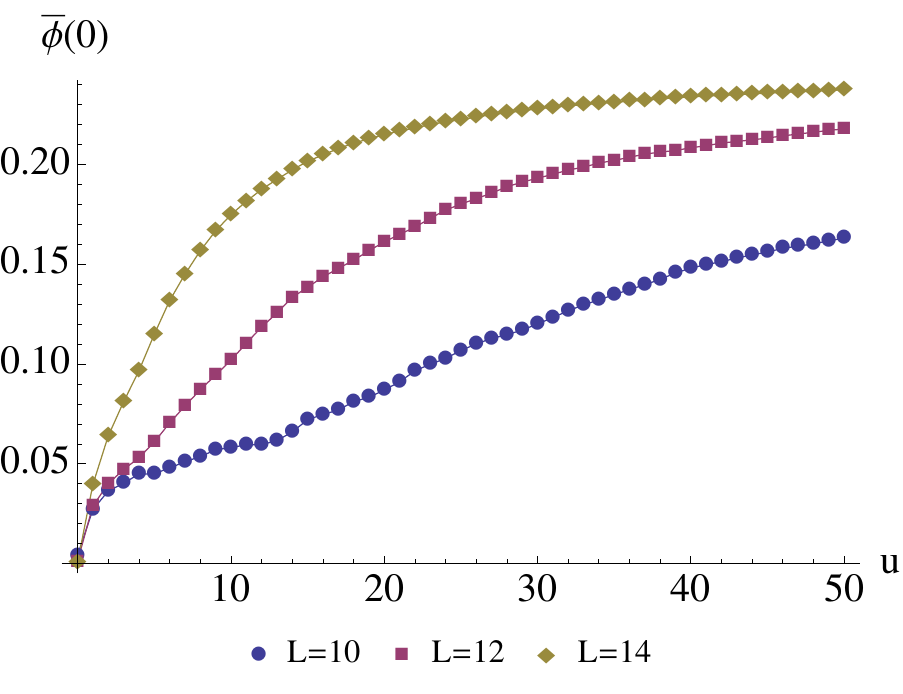}

\caption{Participation ratios for a number of system sizes\label{fig:PR}}
\end{figure}

\begin{figure}
\includegraphics[scale=.8]{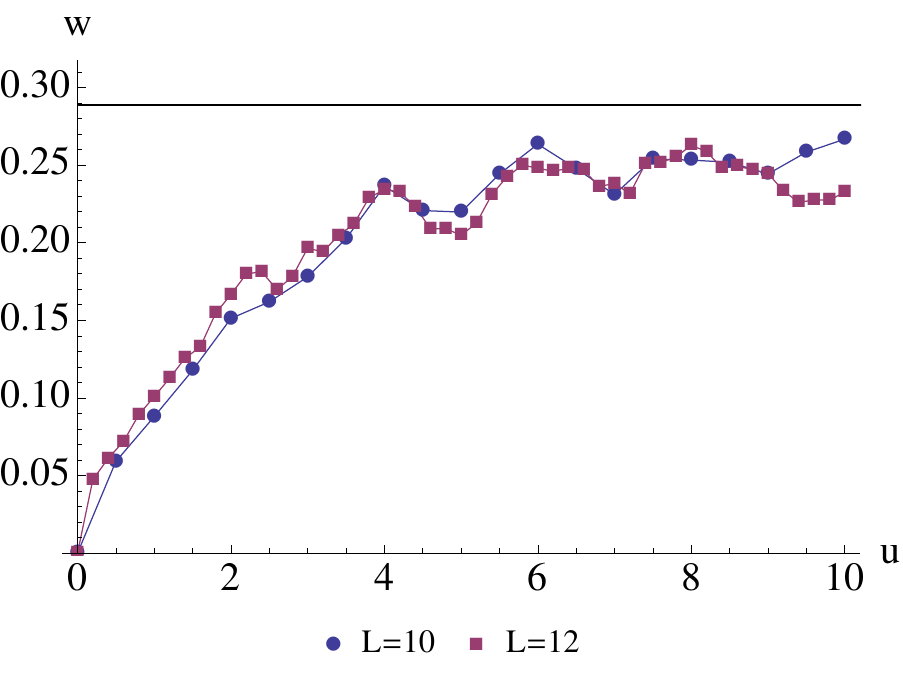}

\caption{Width of bandwidth of $H_{S}$ participating in each eigenstate of
$H_{eff}(0)$, averaged over all its eigenstates for the superlattice Hamiltonian described in the text. 
The solid line indicates
the result for the limit in which all states participate equally.\label{fig:width}}
\end{figure}

Figure \ref{fig:PR} shows the average participation ratios $\phi\left(0\right)$
for a number of system sizes as a function of $u$. Evidently, the
fraction of undriven eigenstates involved in each $\ket{\alpha\left(0\right)}$
is finite. Figure \ref{fig:width} then shows that the participating
states are not concentrated in some region of the spectrum of $H_{S}$
but rather occupy the entire bandwidth (compare the black line which
indicates the result for a uniform distribution throughout the band).

\section{Breathing-trap potential}
Here we show plots of EEVs, dynamics and the final state for a Hamiltonian of the same 
form as in the main text (Eq. (5)) but with a time-dependent potential
\begin{equation}
	V_{i}(t)=\left(\left(i-L/2\right)/\ell_{ho}(t)\right)^2
	\label{eq:breathing-potn}
\end{equation}
with $\ell_{ho}(t)=\ell_0+\widetilde{u}\left(t\right)\delta \ell$ and $\widetilde{u}(t)=+1$ for $0<t<T/2$ and $\widetilde{u}(t)=-1$
for $T/2\leq t\leq T$ (see Supp.~Mat. for another example). We 
take $\ell_0=5$ and $\delta \ell=1$, and, again, $J=V_{1}=V_{2}$.

\begin{figure}
\includegraphics[scale=0.8]{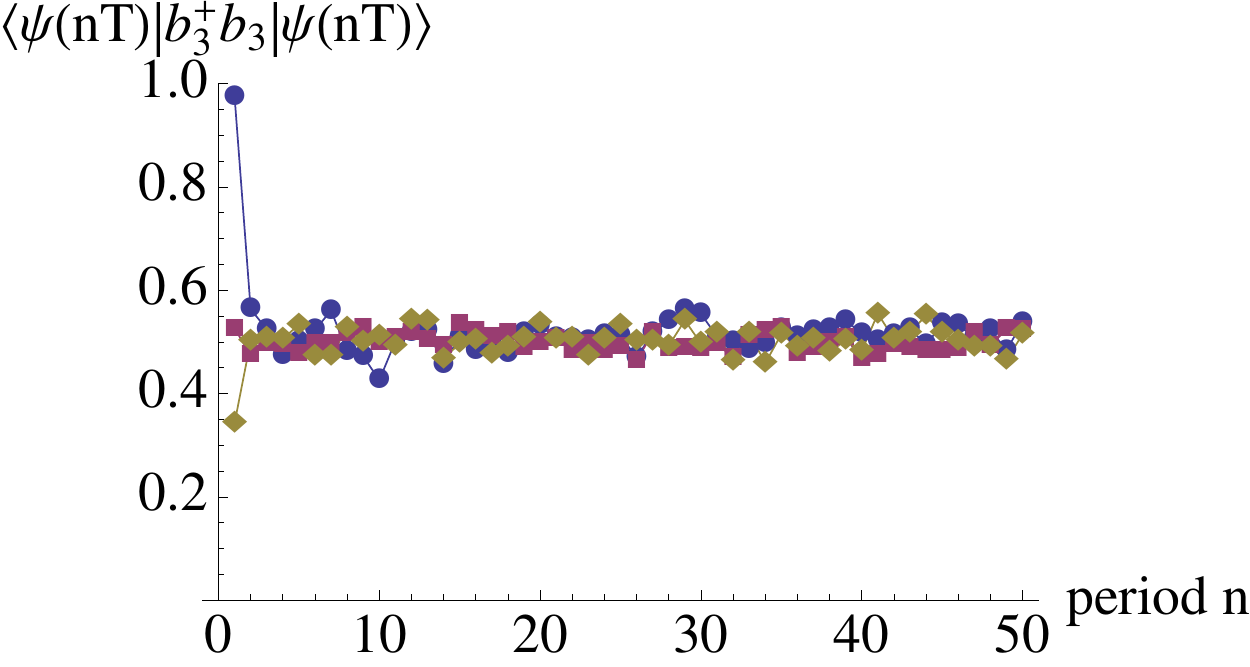}
\caption{Stroboscopic observation of the density at site $i=3$ 
for a breathing trap as a function of period. The initial
states are prepared as described in the main text, and the
observations are made at the beginning of the period,
$\epsilon=0$.
\label{fig:density-breathing-trap}}
\end{figure}

\begin{figure}
\includegraphics[scale=0.8]{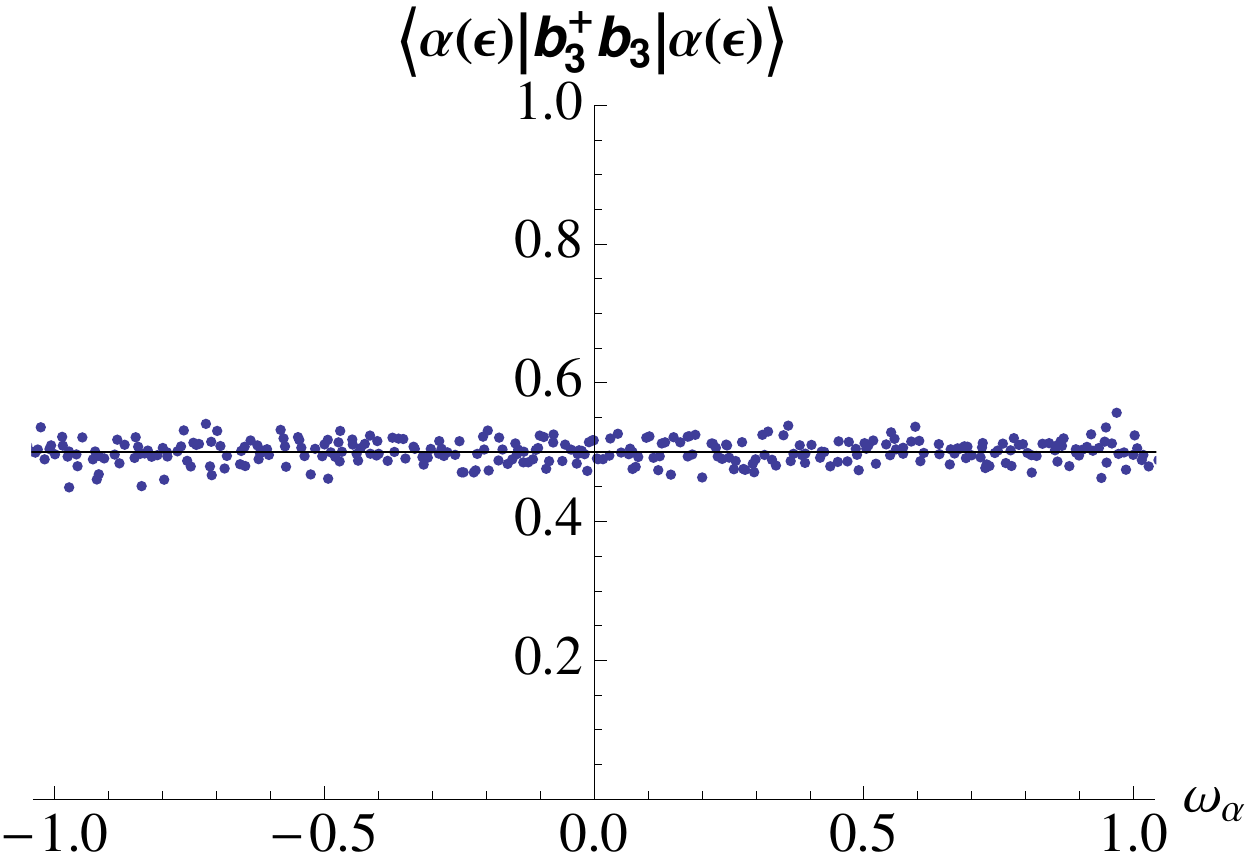}
\caption{Eigenstate expectation values for the density at site $i=3$ for a breathing 
trap for $\epsilon=0$.\label{fig:eev-trap}}
\end{figure}

\begin{figure}
\includegraphics[scale=0.8]{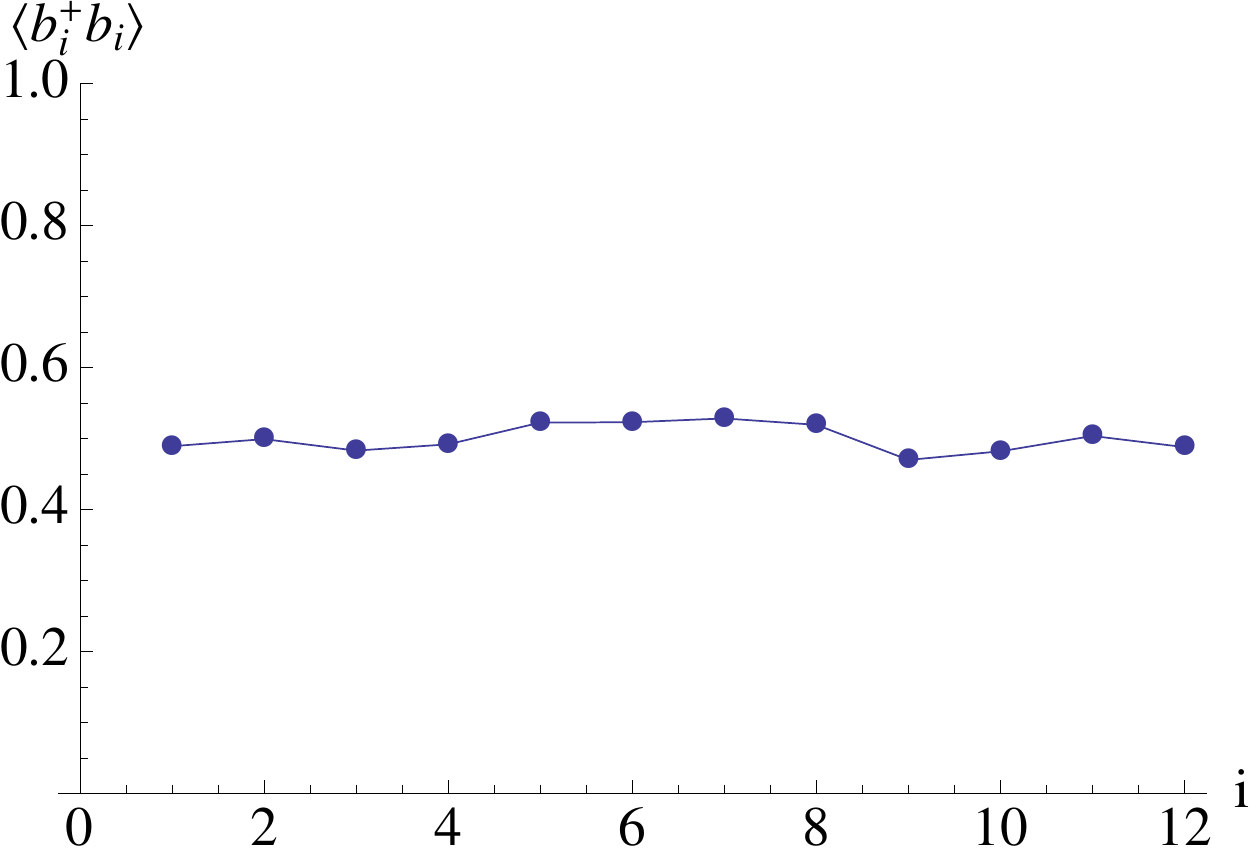}
\caption{Snapshot of final density for a ``breathing trap'' potential, corresponding
to the last point of the blue line of Fig.~\ref{fig:density-breathing-trap}. Note the 
spatially uniform density despite the strong DC component in the quadratic ``trapping'' 
potential. \label{fig:final-density-trap}}.
\end{figure}

Figure~\ref{fig:density-breathing-trap} displays the dynamics evolution for three intial
states selected as described in the main text, then evolved with the Hamiltonian of
Eq. (5) of the main text but with the potential of Eq.~\eqref{eq:breathing-potn},
for system size $l=12$ and $N=6$ particles. Note that, again, all three states evolve 
to the same stationary state. This is understood again from the flatness of the
EEVs, shown in Fig.~\ref{fig:eev-trap}. Finally, a snapshot at long times of one of the 
states is shown in Fig.~\ref{fig:final-density-trap}. Note that, despite the strong
DC component of the potential, the density is spatially uniform. This is as expected, 
since, according to the discussion in the main text, the density at site $i$ is given
by $\mathrm{tr}\left(b^\dagger_i b_i\right)$.

\end{document}